\documentclass[two column,jgrga]{AGUTeX}

  \usepackage[dvips]{graphicx}
  \setkeys{Gin}{draft=false}
\usepackage{pdflscape}
\usepackage{multirow}
\usepackage{array}
\usepackage{color}
\usepackage{amsmath,amssymb}

\newcommand{\red}{}
\newcommand{\blue}{}

\authorrunninghead{Ding, Li, et al}

\titlerunninghead{time interval in the twin-CME scenario}

\begin{document}

%
%

\title{On the identification of time interval threshold in  the twin-CME scenario}
%
%

%
%


%
\authors{Ding, Liu-Guan \altaffilmark{1},
 Li, Gang \altaffilmark{2,*},
 Dong, Li-Hua \altaffilmark{3},
 Jiang, Yong \altaffilmark{4},
 Jian, Yi \altaffilmark{4},
 and Gu,~Bin \altaffilmark{1}}

\altaffiltext{1}{School of Physics and Optoelectronic Engineering, Nanjing University
     of Information Science and Technology, Nanjing 210044, China.}
\altaffiltext{2}{Department of Space Science and CSPAR, University of Alabama in Huntsville,
Huntsville, AL35899, USA;  gang.li@uah.edu.}
\altaffiltext{3}{School of Media Technology, Communication University of China in Nanjing, Nanjing 211172, China.}
\altaffiltext{4}{College of Math and Statistics, Nanjing University of Information Science and Technology,
             Nanjing 210044, China.}


%
%


\begin{abstract}
Recently it has been suggested that the ``twin-CME'' scenario \citep{Li.etal12} may be a very effective
mechanism in causing extreme Solar Energetic Particle (SEP) events and in particular Ground Level Enhancement
(GLE) events. \citet{Ding.etal13} performed a statistical examination of the twin-CME scenario with a
total of $126$ fast and wide western Coronal Mass Ejections (CMEs). They found that CMEs having a
preceding CME with a speed $>$ 300 $km/s$ within $9$ hours from the same active region have larger
probability of leading to large SEP events than CMEs that do not have preceding CMEs.
The choice of $9$ hours being the time lag $\tau$ between the preceding CME and the main CME
was based on some crude estimates of the decay time of the turbulence downstream of the shock driven
by the preceding CME. In this work, we examine this choice. For the $126$ fast wide CMEs examined
in \citep{Ding.etal13}, we vary the time lag $\tau$ from $1$ hour to $24$ hours with an increment of $1$ hour.
By considering three quantities whose values depend on the choice of this time lag $\tau$, we show that the
choice of $13$ hours for $\tau$ is more appropriate.
Our study confirms our earlier result that twin CMEs are more likely to lead to large SEP events than
single fast CMEs. The results shown here are of great relevance to space weather studies.
\end{abstract}

%
%

%

\begin{article}

%
%

\section{Introduction}           
\label{sect:intro}

A major concern of Space Weather is Solar Energetic Particle events (SEPs). In many SEPs, ions can reach
energies  $\sim$ GeV/nuc. It is now largely believed that these energetic particles are accelerated
at/near the Sun via mainly solar flares and/or coronal mass ejections (CMEs).
Historically ``impulsive'' events refer to those events where particles are accelerated at flares  and
``gradual''  events refer to those where particles are accelerated by CME-driven shocks
 \citep{Cane.etal86, Reames95, Reames99}.

Often large SEP events are gradual events and are almost always associated with fast CMEs.
On the other hand, it has been noted for a long time that not all fast CMEs can \red{be associated with} SEP events
measured in the near Earth environment by GOES and/or ACE \citep{Kahler96}.
To explain the observed high intensity of SEPs in large SEP events,
\citet{Kahler.etal00} noted that the ambient energetic
particle intensity prior to the event may be important in causing a large SEP event. The origins of
these seed particles may be preceding impulsive flares \citep{Mason.etal99, Mason.etal00} or
preceding CMEs \citep{Gopalswamy.etal04}.

In an earlier study of $57$ large SEP events that had intensity $> 10$ pfu \blue{(Particle Flux Units,
1~pfu = 1~proton$\rm {/({cm^2}\cdot~s\cdot ~sr)}$)} at $> 10$ MeV,
\citet{Gopalswamy.etal04} showed that there exists a strong correlation between
high particle intensity and preceding CMEs. In the work of \citet{Gopalswamy.etal04}, a preceding CME
occurs within $24$ hours of the primary CME. Following \citet{Gopalswamy.etal04},
\citet{Li.Zank05} suggested that if two CMEs erupt closely in time and both drive shocks, then the
shock driven by the preceding CME can lead to a much enhanced turbulence at the shock driven by the
second CME, therefore lead to a more effective acceleration process and a large SEP event.
Based on a simple estimation of the acceleration time scale, \citet{Li.Zank05}
showed that the maximum particle energy at the second shock can be $\sim 32$ times larger than
that at a single shock case. \citet{Li.Zank05} also noted that the preceding shock can provide
seed particles for the main shock.

Later, in a study of $16$ Ground Level Enhancement (GLE) events in solar cycle $23$,
\citet{Li.etal12} found that there were always preceding CMEs within $9$ hours of the main CME.
This led \citet{Li.etal12} to propose the twin-CME scenario for GLE events. In this scenario,
two CMEs erupt from the same active region (AR) closely in time. The first CME drives a shock,
which, although can accelerate particles, but may not reach energies above $10$ MeV/nucleon.
As particles are accelerated and escape from shock front, they will excite
Alfv\'{e}n waves upstream the shock which get transmitted to downstream of the shock and are enhanced.
When the second CME and its driven shock propagate into this wave-enhanced region, which is also populated
by pre-accelerated ions, more efficient particle acceleration will occur and particles can be accelerated
to very high energies.
In the twin-CME scenario, magnetic reconnection between field lines that
drape the second CME and that enclose the first CME may occur. As such, the material inside
the first CME's driver can be processed by the second CME, leading to a possible enhancement of
heavy ions that are compositionally flare-like \citep{Li.etal12}.


Continue the work of \citet{Li.etal12},  \citet{Ding.etal13} tested the twin-CME scenario against all
large SEP events and fast CMEs with speed $>$ 900 km/s from the western hemisphere
in solar cycle $23$. They found that many single fast CMEs do not lead to large SEP events
and most large SEP events agree with the twin-CME scenario.
They also noted that the peak fluxes in those SEP events that are associated with ``twin CMEs'' show
no correlation with the speed of the associated fast CMEs, nor with the associated flare classes.
To examine if the seed population plays an important role in causing a large SEP event,
\citet{Ding.etal13} examined the daily average low energy seed proton intensity 24 hours prior
to SEP onset time. They found that for single-CME events there seems to exist a correlation between
the low energy seed intensity and the SEP peak flux. For twin-CME events, however, no such correlation
exists.  This can be nicely explained by the twin-CME scenario because the presence of a preceding
CME can effectively lead to an enhanced seed population, therefore no pre-existing seed population
is needed.

The work of \citet{Ding.etal13} was a statistical study. Lately \citet{Shen.etal13} examined
the 17 May 2012 GLE event, the first GLE of solar cycle $24$. Using multiple spacecraft observations,
these authors identified two ejections, with a separation of only $3$ minutes in the 17 May 2012 event.
For such a small separation, not only the enhanced turbulence at the second shock is important in
accelerating particles, the trapping of particles between the two shocks can also
lead to extra contribution to the acceleration process.

According to \citet{Li.etal12}, two CMEs occur within $9$ hours can be regarded as twin CMEs.
If the separation between the two CMEs exceed $9$ hours, they are regarded as single CMEs. The same
interval was also use in \citet{Ding.etal13}. However, the choice of $9$ hours is based on a crude
estimate of the decay time of the turbulence downstream of the first CME-driven shock
\citep{Li.etal12}. Therefore, it is best regarded as a practical choice with ambiguity.
Can we obtain a better choice of the time lag between the two CMEs so that they are
classified as twin CMEs?  We address this question in this work.

Our paper is organized as follows: in section~\ref{sect:data} we discuss the data selection and
 analysis procedure; in section~\ref{sect:result} we present our analysis results;
section ~\ref{sect:disc} contains the discussions and the conclusions.

\section{Data selection}
\label{sect:data}

For the CMEs, we use the list from \citet{Ding.etal13}, which include $126$ fast
and wide CMEs covering the range of 1997 to 2006. The identification criteria of these CMEs
are \citep{Ding.etal13} :
(1) the speed is faster than 900 km/s,(2) the angular width (WD) is larger than $60^\circ$,
and (3) the source location is on the western hemisphere with a longitude smaller than $90^\circ$.
%
%
\red{The definition of a large SEP event can be subjective.
As a working definition, we use the proton event list from the National Oceanic and
Atmospheric Administration (NOAA)  ``Solar Proton Events Affecting the Earth Environment'' list at
http://www.swpc.noaa.gov/ftpdir/indices/SPE.txt
as our large SEP event list. Note that the Space Weather Prediction Center (SWPC) Space Weather Operations (SWO)
defines an event with an intensity $>10$ pfu in  the $>10$ MeV channel of  the
 Geostationary Operational Environmental Satellite (GOES) instrument as a Minor Solar Radiation Storm event. }

In identifying the preceding CMEs, \blue{we follow \citet{Ding.etal13}
and use both EIT observations and LASCO observations. Movies are examined manually on a case
by case basis to identify the AR for the preceding CMEs. For disk events, the identification of AR source
regions (e.g. flares) of these preceding CMEs is relatively easy. For limb events or events with source
region having longitudes $>60^{\circ}$, the identification of the AR can be sometimes tricky. In a future
study we plan to examine large SEP events in solar cycle 24 where observations from STEREOs can
help to remove some of the source region ambiguities.
We also follow \citet{Li.etal12} and \citet{Ding.etal13} and set a speed threshold of $300$ km/s.
 we use the plane of sky speed. Depending on the source AR
longitude and the propagation direction, the projected speed can be somewhat larger.}
Our choice of  $300$ km/s being the threshold is such that the preceding CME is most likely
super-Alfv\'{e}nic, therefore capable of driving a shock \citep{Evans.Opher08}.
Besides the CDAW Data Center catalog at
http://cdaw.gsfc.nasa.gov/CME\_list/, we also make use of another two online catalogs including
the Solar Eruptive Event Detection System (SEEDS) catalog at
http://spaceweather.gmu.edu/seeds/ \citep{Olmedo.etal08}, and the Computer Aided CME Tracking (CACTus)
catalog at http://sidc.oma.be/cactus/ \citep{Robbrecht.Berghmans04}.
We use the CDAW database if a CME exists in multiple databases.

Based on these conditions, we identify all preceding CMEs for any given main CME up to $24$ hrs.
We list the main CMEs and the corresponding preceding CMEs in Table~\ref{table.1}.
The first column in Table~\ref{table.1} is the event number. The second column is the onset time of
the main CME. The third column is the NOAA active region (AR) number, and `fp' denotes events that have
no NOAA AR numbers. Column $4$ is the location of the source region at the time of the main CME erupting.
Column $5-14$ are the onset time of each preceding CME of main CME within 24hrs. The column labelled
`p1' lists the closest preceding CMEs ahead the main CMEs; similarly the `p2' column lists the second
closest preceding CMEs, and the `p3' column lists the third closest preceding CMEs, and so on.  In these
columns symbol `-' denotes that no preceding CMEs are identified. The `*' symbol in column 15 denotes
that a large SEP event is produced.

\section{Analyses and Results}
\label{sect:result}

For our analysis, we follow \citet{Ding.etal13} and
 categorize all CMEs in our study to four groups. Group I are ``twin CMEs'' that lead to large SEPs;
group II are single CMEs that lead to large SEPs; group III are
``twin CMEs'' that do not lead to large SEPs; and group IV are single CMEs that do not lead to large SEPs.
We use $N_i(\tau)$ where $i=$I, II, III, IV to denote the number of events in each of these $4$ groups.
Clearly, $N_{i}$ depends on the choice of the time lag $\tau$ between the preceding CME and the main CME.

%
%

\begin{figure}[htb]
\centering
\noindent\includegraphics[width=0.8\textwidth]{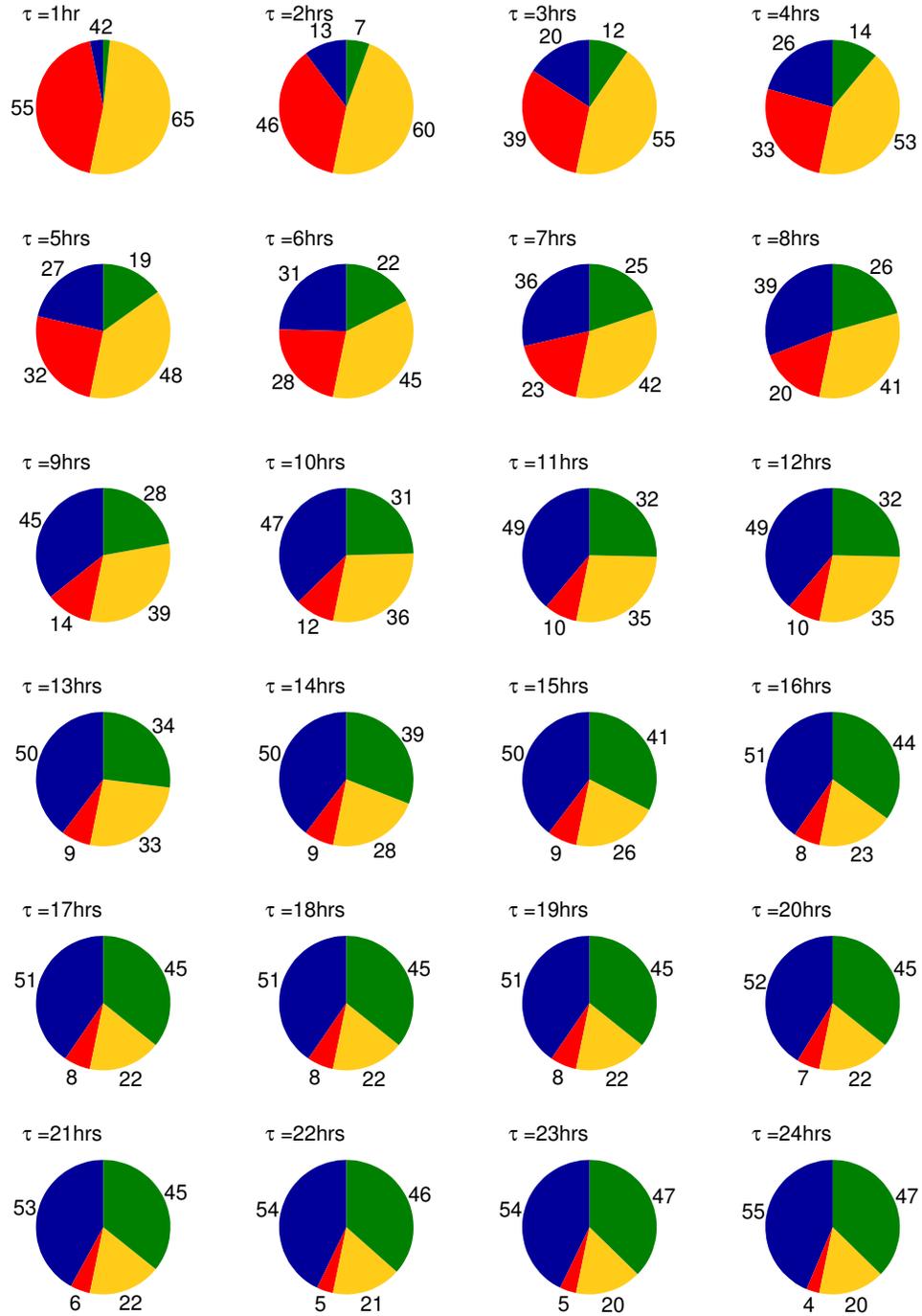}
   \caption{The statistic pie chart of preceding CMEs in twin-CME scenario using different time threshold $\tau$ ahead of 126
fast main CMEs in solar cycle 23. The value of $\tau$ are shown on top of each chart.
 The blue section indicates ``twin CMEs'' that generate SEP events (Group I), the green section indicates the ``twin CMEs''
that fail to produce SEP events (Group III), the red section indicates ``single CMEs'' that generate SEP events (Group II),
and the yellow section indicates ``single CMEs'' that fail to   produce SEP events \red{(Group IV)}.
   }
   \label{Fig:pieplot}
\end{figure}

Figure~\ref{Fig:pieplot} shows the statistic pie chart of the number of events in all four groups
as $\tau$ increases from 1~hr to 24~hrs. In Figure~\ref{Fig:pieplot}, $N_{\rm I}$ is shown as blue,
$N_{\rm II}$  red, $N_{\rm III}$ green, and $N_{\rm IV}$ yellow. The first row is for $\tau=1$,2,3,4 hrs,
the second row is for $\tau=5$,6,7,8 hrs, etc.

To examine how good the choice of $9$ hrs is, we consider the following quantities which are formed
from the number of events in each group:

\begin{eqnarray}
r_1 &=& \frac{N_{\rm I}}{N_{\rm I}+N_{\rm III}}  \label{eq:r1} \\
r_2 &=& \frac{N_{\rm I}}{N_{\rm I}+N_{\rm II}}  \label{eq:r2} \\
r_3 &=& \frac{N_{\rm IV}}{N_{\rm II}+N_{\rm IV}}  \label{eq:r3}\\
\red{r_4} &\red{=}& \red{\frac{N_{\rm I}+N_{\rm III}} {N_{\rm I}+N_{\rm II}+N_{\rm III}+N_{\rm IV}}}   \label{eq:r4}
\end{eqnarray}

So $r_1$ is the ratio of number of twin CMEs that lead to large SEPs to the total number of twin CMEs;
 $r_2$ is the ratio of number of twin CMEs that lead to large SEPs to the total number of large SEPs;
 $r_3$ is the ratio of number of single CMEs that do not lead to large SEPs to the total number of single CMEs;
\red{and $r_4$ is the ratio of the total number of twin CMEs to the total number of CMEs. Note that
$r_1$ is the same of $r_2/r_4$, up to a constant. As we explain below, we use $r_4$ as an auxiliary quantity to
remove the general $\tau$ dependence of $r_2$.}

We now vary the time interval $\tau$ from $1$ hr to $24$ hrs with a one-hour time step and examine how $r_i$ vary
with $\tau$. We assume that twin CMEs are much more efficient than single CMEs in accelerating particles
and most of the large SEP events are caused by twin CMEs. Our goal is to identify the time lag $\tau^{*}$ that
defines two CMEs as twin CMEs. Note, as discussed in \citet{Li.etal12}, $\tau^{*}$
is physically related to the decay time scale of the turbulence downstream of the first shock.
We assume that \red{the effect of the first CME on particle acceleration at the shock driven by the second CME
can be ignored} when two CMEs from the same AR are separated by a time $\tau>\tau^{*}$.

Clearly as $\tau$ increases, the number of twin CMEs that lead to large SEPs (i.e. $N_{\rm I}$) increases;
at the same time, the total number of events that can be categorized as twin CMEs (i.e. $N_{\rm I}+N_{\rm III}$) also increases.
When $\tau$ is larger than $\tau^{*}$, however, the effect of the preceding CME on the main CME becomes negligible,
then as $\tau$ further increases, $N_{\rm I}$ remains almost unchanged but $N_{III}$ keeps increasing, therefore we expect
to see a drop of $r_1(\tau)$ at $\tau^{*}$.

Next consider $r_2(\tau)$. First note that the denominator of $r_2(\tau)$, which is the total number of SEPs, is a constant;
the numerator, $N_{\rm I}$, increases as $\tau$ increases. Initially when $\tau<\tau^{*}$,
the increase rate of $N_{\rm I}$ is fast. If all large SEPs are caused by twin CMEs, then when $\tau>\tau^{*}$,
$N_{\rm I}$ stops increasing and becomes a constant. Of course, not all large SEPs are caused by twin CMEs and
some single CMEs can also lead to large SEPs, so $N_{\rm I}$ still increases when $\tau>\tau^{*}$, but will be at a much
smaller rate than when {$\tau<\tau^{*}$}. Consequently, $r_2(\tau)$ will increase with $\tau$, but the increase rate
will show a clear drop at $\sim \tau^{*}$. \red{
Note that one may wonder if the dependence of $r_2$ on $\tau$ may be due to simply the fact
that the total number of twin CMEs increases with $\tau$. To examine that, an auxiliary ratio $r_4$ is also considered.
This is the total number of twin CMEs (i. e. $N_{\rm I}$ and $N_{\rm III}$) to the total number of CMEs.
Since when $\tau<\tau^{*}$, $N_{\rm I}$ increases as $\tau$ increases, so $N_{\rm I} + N_{\rm III}$ also increases with $\tau$.
We therefore expect the general trend of $r_4$ is similar to $r_2$ when $\tau<\tau^{*}$. However, when $\tau>\tau^{*}$,
 $N_{\rm I}$ stops increasing with $\tau$,  but  $N_{\rm III}$, therefore $N_{\rm I} + N_{\rm III}$ still increases with $\tau$.
So the similar trend of $r_2$ and $r_4$ will stop at $\tau>\tau^{*}$. }

Finally consider $r_3(\tau)$.  As $\tau$ increases, both the number of single CMEs not leading to SEPs (i.e. $N_{\rm IV}$)
and the total number of single CMEs (i.e. $N_{\rm II}+N_{\rm IV}$) decreases, but the latter decreases faster
(since a lot of single fast CMEs do not lead to large SEPs). Therefore $r_{3}(\tau)$ increases
with $\tau$ until $\tau^{*}$, after which a decrement of $N_{\rm IV}$ will likely lead to a decrement $N_{\rm II}$,
so  we expect the increase rate of $r_3(\tau)$ (i.e. the slope) to also drop around $\tau^{*}$, similar to $r_2(\tau)$.

\begin{figure}[htb]
\centering
\noindent\includegraphics[width=0.5\textwidth]{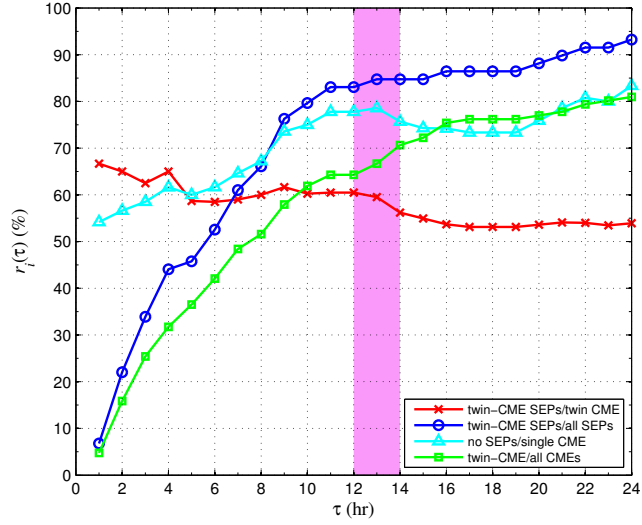}
   \caption{ Plots of $r_1$, $r_2$, $r_3$, \red{and $r_4$} as defined in equations~(\ref{eq:r1}) to \red{(\ref{eq:r4})}.
The figure shows the percentage of events in each of the four different groups as a function of
the time interval $\tau$ between the two CMEs.
The red line with crosses is $r_1$; the blue line with circles is $r_2$; the cyan line with triangle is $r_3$; \red{and the green line with squares is $r_4$}.
}
   \label{Fig:percentage}
\end{figure}

Figure~\ref{Fig:percentage} plots $r_1$ (red), $r_2$ (blue),  $r_3$ (cyan), \red{and $r_4$ (green)}.
\red{As we explained above, the best indicator of $\tau^{*}$ is $r_1$. This is because the variation of $r_1$ is very
gradual, so even a change of $\sim 10\%$ can be clearly identified. }
From the figure we can see that there is a noticeable drop of $r_1$ at $\tau \sim 13$ hrs.
Furthermore, the slope of $r_2$ and $r_3$ show significant decreases at \red{ $\tau \sim 12$} hrs.
\red{Note that although $r_4$ increases with $\tau$ in a similar manner as $r_2$ when $\tau<13$ hrs (as it should), after
$\tau>13$ hrs, it keeps increasing with the same rate until  $\tau \sim 17$hrs while $r_2$ becomes roughly a constant
between $13<\tau<17$ hrs.  The $\tau$-dependence of the ratios of $r_1$ to $r_4$ agree nicely to the prediction of the
twin-CME scenario. }

\red{ Figure~\ref{Fig:percentage} suggests a reasonable choice of $\tau^{*}$ for the twin-CME scenario
is between $12$ and $13$ hours. Since  $r_1$ is the most sensible quantity to $\tau^{*}$, we  set
$ \tau^* =13 \pm 1$ hrs in the following. }

Some useful knowledge concerning large SEPs can be immediately read off from
Figure~\ref{Fig:pieplot} and Figure~\ref{Fig:percentage}.
For example, with $\tau^{*}=13$ hrs as the criteria for twin CMEs, we find that $\sim 60\%$ twin CMEs lead to
large SEPs. In comparison, only $\sim 21\%$ single CMEs lead to large SEPs.
 Furthermore, we also find that, from Figure~\ref{Fig:pieplot}, the percentage of twin CMEs in large SEP events
is $\sim 85\%$, which is significantly higher than $\sim 67\%$, the percentage of twin CMEs in all
fast and wide CMEs. These conclusions are particularly useful for space weather forecasting.

\begin{figure}[htb]
    \centering
    \noindent \includegraphics[width=0.5\textwidth]{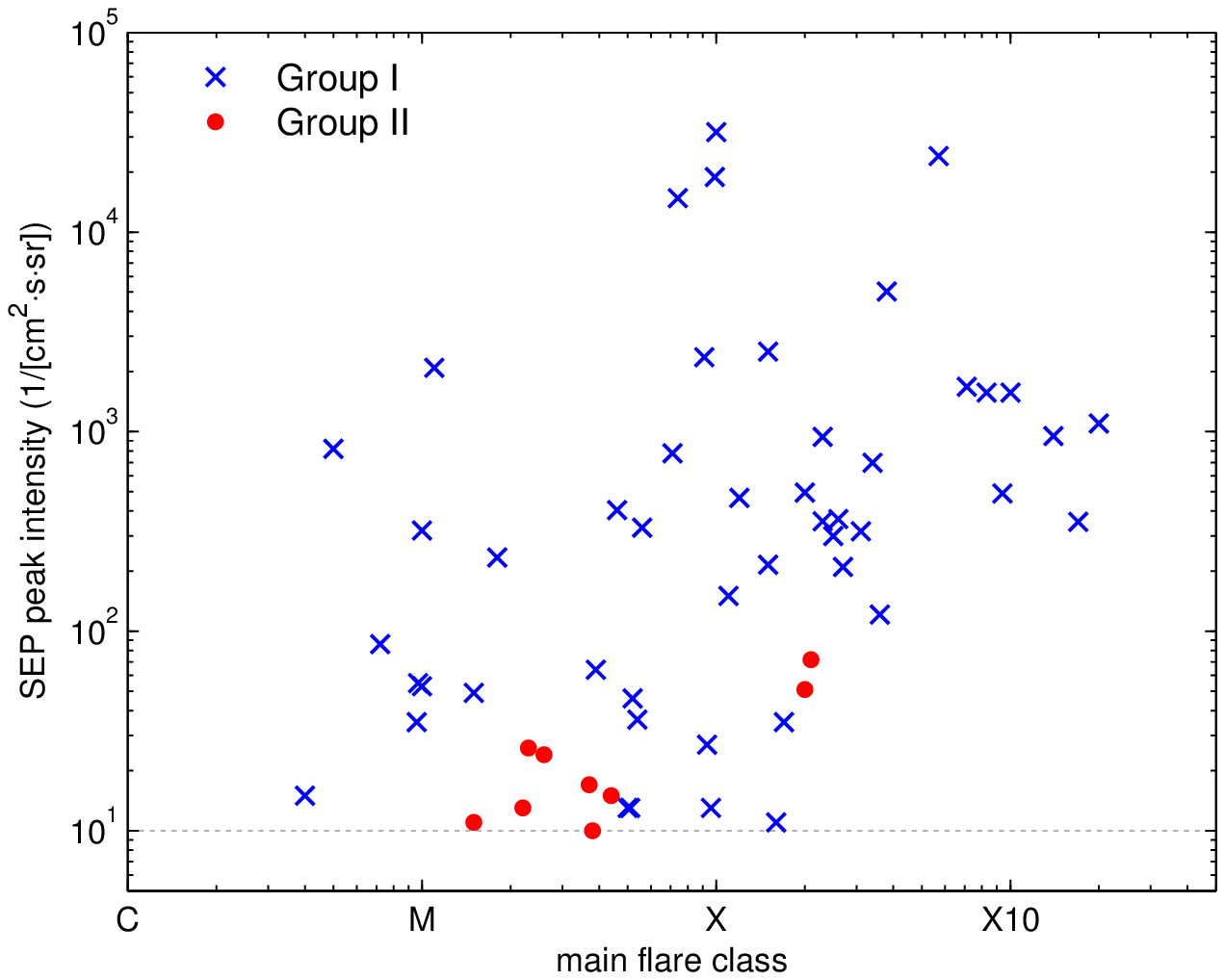}
    \caption{The peak flux of all $59$ western SEP events from Table~\ref{table.1}. The $x$-axis is the flare class.
The $y$-axis is the $>$ 10 MeV peak proton flux measured by the GOES spacecraft. Twin CME events using \red{$\tau=13$} hrs
are labeled as blue crosses (group I) and single CME events are labeled as red dots (group II).  }
    \label{fig.corr_flare}
\end{figure}

\begin{figure}[htb]
    \centering
    \noindent \includegraphics[width=0.5\textwidth]{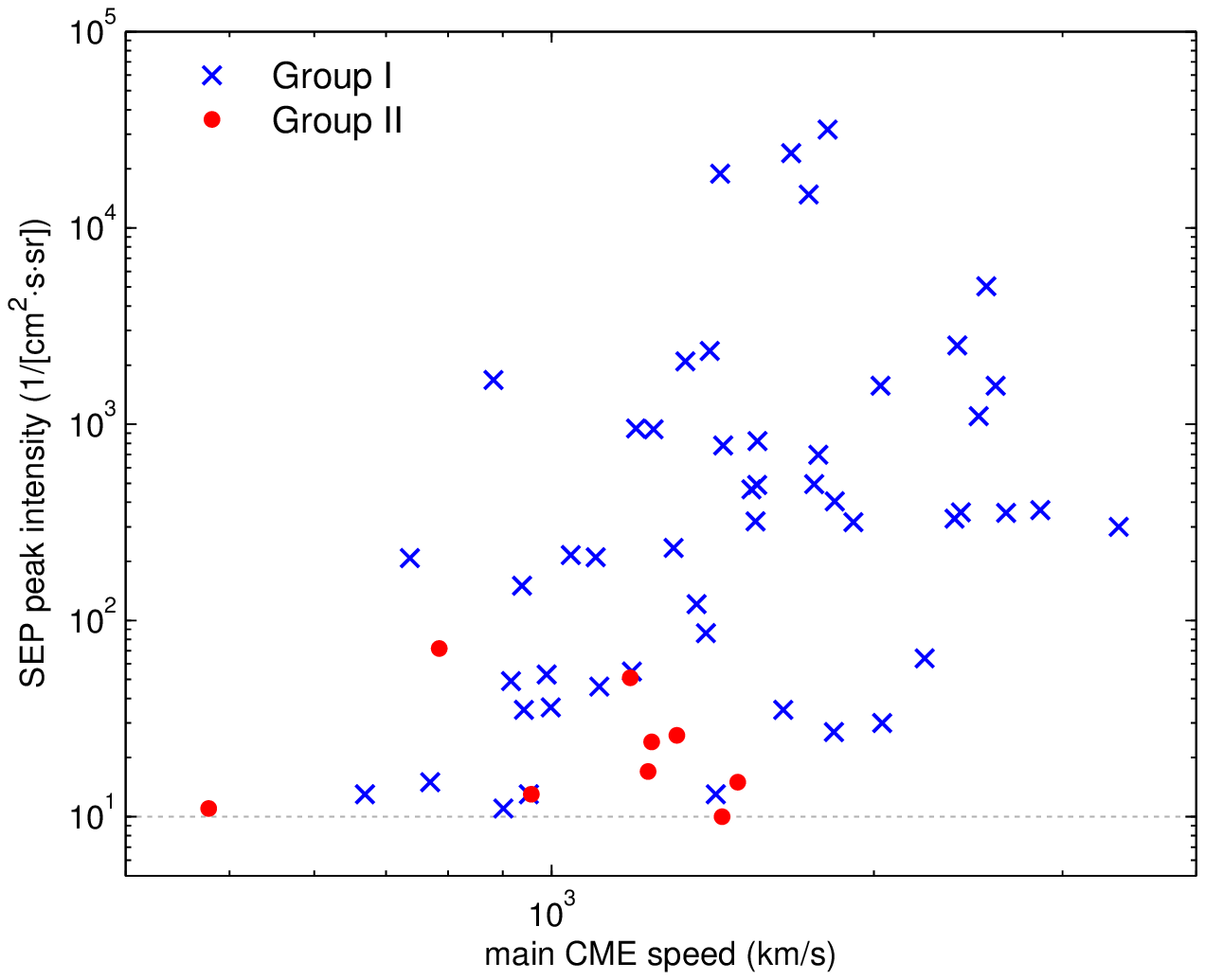}
    \caption{The peak flux of all $59$ western SEP events from Table~\ref{table.1}. The $x$-axis is the CME speed.
The $y$-axis is the $>$ 10 MeV peak proton flux measured by the GOES spacecraft. Twin CME events using \red{$\tau=13$} hrs
are labeled as blue crosses (group I) and single CME events are labeled as red dots (group II).  }
    \label{fig.corr_cme}
\end{figure}

\begin{figure}[htb]
    \centering
    \noindent \includegraphics[width=0.5\textwidth]{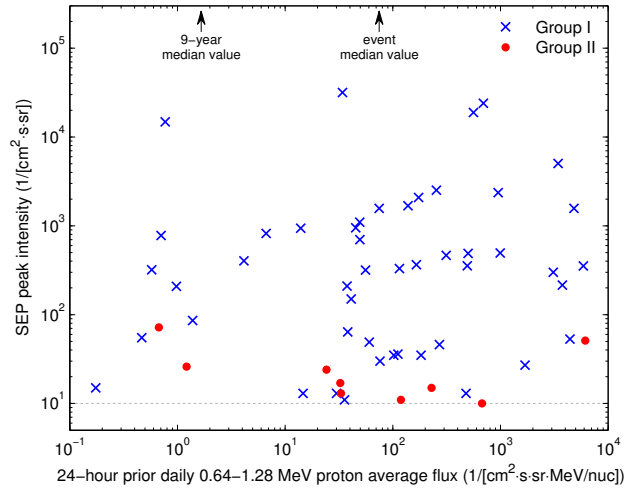}
    \caption{The peak flux of all $59$ western SEP events from Table~\ref{table.1}. The $x$-axis is the the
1-day prior daily average proton intensity from the  Ultra Low Energy Isotope Spectrometer (ULEIS)
instrument ($0.64$ to $1.28$ MeV channel) on board the Advanced Composition Explorer (ACE) spacecraft.
The $y$-axis is the $>$ 10 MeV peak proton flux measured by the GOES spacecraft. Twin CME events using \red{$\tau=13$} hrs
are labeled as blue crosses (group I) and single CME events are labeled as red dots (group II).
 The arrows indicate the median values of 1-day
prior daily average proton intensity of all $59$ events and of daily average proton intensity from 1998 to 2006 in
$0.64-1.28$ MeV channel. }
    \label{fig.corr_seed}
\end{figure}

In \citet{Ding.etal13}, we examined the correlation between the peak intensity and the flare class, the main CME speed,
and the seed population  for both twin CMEs and single CMEs, assuming $\tau^{*}=9$ hrs.
Now with the choice of \red{$\tau^{*}=13$} hrs, these are replotted in Figure~\ref{fig.corr_flare},
Figure~\ref{fig.corr_cme} and Figure~\ref{fig.corr_seed}.

In all three figures, the red dots are SEP events caused by single CMEs and the blue crosses are SEP events caused by
twin CMEs. There were $9$ SEPs which were caused by single CMEs, all of them have
their peak intensities below $100$ pfu.


From these figures we see that for both single CME and twin CME events,
neither the flare class, nor the CME speed, nor the 24-hour prior ACE/ULEIS
\red{(Ultra Low Energy Isotope Spectrometer, ULEIS, \citep{Stone.etal98})} ion measurements correlates well with the
peak intensity of the event. Note however, that the ULEIS measurement at 1AU, which we use here as a proxy of the
seed population, does not necessarily reflect the pre-event plasma environment close to the sun.
Our results support the previous work by \citet{Gopalswamy.etal04} who concluded that flare class and CME speeds are not
good indicators of event magnitude.

There are $34$ events with peak intensity $> 100$ pfu. All of them are twin-CME events.
This is an important observation. It clearly shows that event \red{magnitude} has a strong
dependence on the presence of a preceding CME.

\section{Conclusions and Discussions}
\label{sect:disc}

In this work, we extend the work of \citep{Ding.etal13} and further examined the twin-CME scenario.
In particular, we examined what is the best value of the time interval threshold $\tau$ in identifying a
twin-CME event. For all fast and wide CMEs and large SEPs identified in \citet{Ding.etal13},  we vary $\tau$ from
$1$ to $24$ hours and found that the best value of $\tau$ is \red{$\sim 13$} hours.  As proposed in \citet{Li.Zank05},
if a preceding CME (and its driven shock) occurs within \red{$13$} hrs of a main CME, it can provide both an enhanced
turbulence and enhanced seed population at the second shock, leading to a more efficient acceleration.
In \citet{Li.etal12} the value of $\tau$
corresponds to the decay time of the turbulence downstream the first CME shock \citep{Li.etal12}  and was estimated to be
$9$ hrs.  Our current work improves this estimate. Note that for different events the characteristics (such as
the shock speed, the CME width, etc) of the preceding CMEs can vary largely. The decay time of the turbulence, however,
depends on the Alfv\'{e}n wave speed downstream of the first shock, which is decided by the characteristics of the
solar wind, not the shock,  we therefore expect that the result of $\tau=13$ hrs is not strongly event dependent.

With the choice of $\tau=13$ hrs, the percentage of twin CMEs in all fast and wide CMEs is $\sim 67\%$.
In comparison, the percentage of twin CMEs in large SEP events is $\sim 85\%$, which is $18\%$ higher.
Furthermore, $\sim 60\%$ twin CMEs lead to large SEPs while only $\sim 21\%$ single CMEs lead to large SEPs. Finally,
all large SEP events recorded by GOES with a peak intensity larger than $100$ pfu at $>10$ MeV/nucleon are twin CMEs.
These results suggest that the twin-CME scenario can lead to an efficient particle acceleration process and our findings may
provide a useful basis for space weather forecasting.

In the twin-CME scenario, the role of the preceding CME is to provide the seed population
and the enhanced turbulence \citep{Li.Zank05}. It is tempting to ask which one is more important, the seed population or
the enhanced turbulence? In a series of papers, \citet{Zank.etal00, Li.etal03, Rice.etal03, Li.etal05, Li.etal09,
Verkhoglyadova.etal09, Verkhoglyadova.etal10} have examined particle acceleration at a shock driven by a single
fast CME. They assumed a quasi-parallel shock configuration and an injection efficiency of $\sim 1\%$ and calculated the
enhanced upstream Alfv\'{e}n waves and energetic particle spectrum at the same time using a self-consistent approach.
For strong shocks with a compression ratio close to $4$, they found a maximum proton energy reaching $>500$ MeV.
Their calculations suggest that, at least at quasi-parallel shocks, strong turbulence can be self-generated if there is
enough seed population. Based on these works, the presence of an enhanced seed population at the second shock is probably
more important than the presence of an enhanced wave turbulence at the second shock in causing a large SEP event.
However, we point out that without the enhanced turbulence downstream the first shock, the seed population may
quickly convect or diffuse out from the shock region.

\begin{acknowledgments}
We are grateful to SOHO/LASCO, SOHO/EIT, and CDAW, SEEDS, CACTus CME catalogs for
making their data available online. This work is supported at UAH by NSF grants
ATM-0847719 and AGS-1135432; at NUIST by NSFC-41304150,
NSFC-41174165, NSF for colleges and universities in Jiangsu Province-12KJB170008,
and NUIST Pre-research funding 2013 for LGD.
\end{acknowledgments}

\end{article}

%
%
%
%

\begin{landscape}
\begin{table}[!htbp]
\caption{The onset times of the corresponding preceding CMEs within 24 hours of main CMEs.}
\label{table.1}
\linespread{1.1}
\setlength{\tabcolsep}{1.1pt}
\tiny
\centering
\begin{tabular}{ccccccccccccccc}
  \hline
  \hline
\multirow{2}{*}{No.}&\multirow{2}{1.8cm}{\centering{Onset time of main CME$^a$}} &
\multirow{2}{*}{AR$^b$}&\multirow{2}{*}{Loc.$^c$}&\multicolumn{10}{c}{Onset time of the preceding CMEs$^d$}&
\multirow{2}{*}{Comm.$^e$}\\
\cline{5-14}
&&&&p1&p2&p3&p4&p5&p6&p7&p8&p9&p10&\\
\hline
(1)&(2)&(3)&(4)&(5)&(6)&(7)&(8)&(9)&(10)&(11)&(12)&(13)&(14)&(15)\\
\hline
1&	1997/11/04 06:10&	08100&	S14W33&	1997/11/03 14:26&	1997/11/03 11:11&	1997/11/03 09:53&	--&	--&	--&	--&	--&	--&	--&	*\\
2&	1997/11/06 12:10&	08100&	S18W63&	1997/11/06 04:20&	1997/11/05 12:10&	--&	--&	--&	--&	--&	--&	--&	--&	*\\
3&	1998/01/03 09:42&	08126&	N20W64&	1998/01/02 23:57&	1998/01/02 14:30&	--&	--&	--&	--&	--&	--&	--&	--&	\\
4&	1998/05/02 14:06&	08210&	S15W15&	1998/05/02 05:31&	1998/05/01 23:40&	--&	--&	--&	--&	--&	--&	--&	--&	*\\
5&	1998/05/06 08:29&	08210&	S11W65&	1998/05/06 00:02&	--&	--&	--&	--&	--&	--&	--&	--&	--&	*\\
6&	1998/06/05 07:01&	fp&	S45W20&	1998/06/05 01:30&	--&	--&	--&	--&	--&	--&	--&	--&	--&	\\
7&	1999/06/04 07:26&	08552&	N17W69&	1999/06/04 00:50&	1999/06/03 16:26&	1999/06/03 13:26&	1999/06/03 09:26&	1999/06/03 08:26&	--&	--&	--&	--&	--&	*\\
8&	1999/06/27 09:06&	08592&	N23W25&	1999/06/26 19:54&	1999/06/26 15:54&	--&	--&	--&	--&	--&	--&	--&	--&	\\
9&	1999/06/28 21:30&	08592&	N22W46&	1999/06/28 12:06&	1999/06/28 11:30&	1999/06/28 07:31&	1999/06/28 03:30&	--&	--&	--&	--&	--&	--&	\\
10&	1999/07/25 13:31&	08639&	N38W81&	1999/07/24 22:30&	--&	--&	--&	--&	--&	--&	--&	--&	--&	\\
11&	1999/08/28 01:26&	08674&	S24W29&	--&	--&	--&	--&	--&	--&	--&	--&	--&	--&	\\
12&	1999/09/09 21:54&	08682&	S19W61&	1999/09/09 19:52&	1999/09/09 07:31&	--&	--&	--&	--&	--&	--&	--&	--&	\\
13&	1999/09/19 17:18&	08699&	S21W71&	1999/09/19 10:54&	--&	--&	--&	--&	--&	--&	--&	--&	--&	\\
14&	1999/09/23 15:54&	fp&	S25W50&	--&	--&	--&	--&	--&	--&	--&	--&	--&	--&	\\
15&	1999/11/16 06:54&	08759&	N11W46&	1999/11/16 05:30&	1999/11/15 14:30&	--&	--&	--&	--&	--&	--&	--&	--&	\\
16&	2000/01/05 01:53&	08816&	N23W28&	2000/01/04 11:30&	2000/01/04 08:06&	2000/01/04 06:54&	--&	--&	--&	--&	--&	--&	--&	\\
17&	2000/01/28 20:12&	08841&	S31W17&	--&	--&	--&	--&	--&	--&	--&	--&	--&	--&	\\
18&	2000/02/12 04:30&	08858&	N24W38&	2000/02/11 21:30&	2000/02/11 17:54&	2000/02/11 04:54&	--&	--&	--&	--&	--&	--&	--&	\\
19&	2000/03/04 09:54&	08889&	N20W46&	2000/03/04 03:57&	2000/03/03 18:54&	2000/03/03 16:54&	--&	--&	--&	--&	--&	--&	--&	\\
20&	2000/04/04 16:32&	08933&	N16W66&	2000/04/04 15:06&	--&	--&	--&	--&	--&	--&	--&	--&	--&	*\\
21&	2000/05/05 15:50&	08976&	S12W88&	2000/05/05 03:06&	2000/05/04 21:26&	2000/05/04 16:26&	--&	--&	--&	--&	--&	--&	--&	\\
22&	2000/05/15 16:26&	08993&	S24W67&	2000/05/15 14:26&	2000/05/15 03:06&	--&	--&	--&	--&	--&	--&	--&	--&\\
23&	2000/06/10 17:08&	09026&	N22W38&	2000/06/10 10:17&2000/06/10 08:06&	--&	--&	--&	--&	--&	--&	--&	--&	*\\
24&	2000/06/15 20:06&	09041&	S20W65&	2000/06/15 07:50&	2000/06/15 05:26&	--&	--&	--&	--&	--&	--&	--&	--&	\\
25&	2000/06/25 07:54&	09046&	N16W55&	2000/06/25 07:54&	2000/06/25 06:06&	--&	--&	--&	--&	--&	--&	--&	--&	\\
26&	2000/06/28 19:31&	fp&	N24W85&	--&	--&	--&	--&	--&	--&	--&	--&	--&	--&	\\
27&	2000/07/14 10:54&	09077&	N22W07&	2000/07/14 08:54&	2000/07/14 00:54&	2000/07/13 20:30&	2000/07/13 11:54&	2000/07/13 10:54&	--&	--&	--&	--&	--&	*\\
28&	2000/07/22 11:54&	09085&	N14W56&	2000/07/21 12:54&	--&	--&	--&	--&	--&	--&	--&	--&	--&	*\\
29&	2000/09/12 11:54&	09163&	S17W09&	2000/09/12 10:54&	2000/09/12 08:06&	--&	--&	--&	--&	--&	--&	--&	--&	*\\
30&	2000/09/16 05:18&	09165&	N15W07&	2000/09/15 15:26&	2000/09/15 12:06&	--&	--&	--&	--&	--&	--&	--&	--&	\\
31&	2000/09/16 13:50&	09165&	N14W13&	2000/09/16 09:26&	2000/09/16 08:06&	2000/09/16 05:18&	2000/09/15 15:26&	--&	--&	--&	--&	--&	--&	\\
32&	2000/10/25 08:26&	09199&	N10W66&	2000/10/25 08:06&	2000/10/25 04:06&	2000/10/25 02:50&	--&	--&	--&	--&	--&	--&	--&	*\\
33&	2000/10/26 11:50&	09199&	N17W77&	2000/10/26 07:50&	2000/10/26 01:50&	2000/10/25 23:26&	2000/10/25 16:26&	--&	--&	--&	--&	--&	--&	\\
34&	2000/11/08 23:06&	09213&	N10W77&	2000/11/08 17:06&	2000/11/08 10:26&	2000/11/08 09:26&	2000/11/08 04:50&	2000/11/08 02:06&	2000/11/08 01:27&	--&	--&	--&	--&	*\\
35&	2000/11/24 15:30&	09236&	N22W07&	2000/11/24 11:54&	2000/11/24 11:06&	2000/11/24 07:54&	2000/11/24 05:30&	2000/11/24 04:30&	2000/11/24 02:30&	2000/11/23 23:54&	2000/11/23 21:30&	2000/11/23 18:54&	2000/11/23 17:30&	*\\
36&	2000/11/24 22:06&	09236&	N21W14&	2000/11/24 17:35&	2000/11/24 15:30&	2000/11/24 11:54&	2000/11/24 11:06&	2000/11/24 07:54&	2000/11/24 05:30&	2000/11/24 04:30&	2000/11/24 02:30&	2000/11/23 23:54&--&	\\
37&	2001/01/14 06:30&	fp&	N60W10&	--&	--&	--&	--&	--&	--&	--&	--&	--&	--&	\\
38&	2001/01/26 12:06&	09320&	S23W57&	--&	--&	--&	--&	--&	--&	--&	--&	--&	--&	\\
39&	2001/01/28 15:54&	09313&	S04W59&	2001/01/28 09:30&	--&	--&	--&	--&	--&	--&	--&	--&	--&	*\\
40&	2001/02/11 01:31&	09334&	N24W57&	2001/02/10 09:30&	--&	--&	--&	--&	--&	--&	--&	--&	--&	\\
41&	2001/03/29 10:26&	09393&	N20W19&	2001/03/28 23:50&	2001/03/28 21:50&	2001/03/28 16:26&	--&	--&	--&	--&	--&	--&	--&	*\\
42&	2001/04/02 11:26&	09393&	N17W60&	2001/04/02 05:50&	2001/04/01 23:26&	--&	--&	--&	--&	--&	--&	--&	--&	\\
43&	2001/04/02 22:06&	09393&	N18W82&	2001/04/02 12:50&	2001/04/02 11:26&	2001/04/02 05:50&	2001/04/01 23:26&	--&	--&	--&	--&	--&	--&	*\\
44&	2001/04/05 09:06&	09401&	N24W83&	2001/04/05 02:06&	2001/04/04 16:50&	2001/04/04 10:50&	--&	--&	--&	--&	--&	--&	--&	\\
45&	2001/04/09 15:54&	09415&	S21W04&	2001/04/09 04:54&	2001/04/09 00:06&	--&	--&	--&	--&	--&	--&	--&	--&	\\
46&	2001/04/10 05:30&	09415&	S23W09&	2001/04/10 02:06&	2001/04/09 20:06&	2001/04/09 15:54&	--&	--&	--&	--&	--&	--&	--&	*\\
47&	2001/04/11 00:54&	09417&	S08W32&	--&	--&	--&	--&	--&	--&	--&	--&	--&	--&	\\
48&	2001/04/11 13:31&	09415&	S23W32&	2001/04/10 21:54&	--&	--&	--&	--&	--&	--&	--&	--&	--&	\\
49&	2001/04/12 10:31&	09415&	N19W42&	2001/04/11 13:31&	--&	--&	--&	--&	--&	--&	--&	--&	--&	*\\
50&	2001/04/15 14:06&	09415&	N20W85&	2001/04/15 11:18&	2001/04/15 09:30&	2001/04/15 00:30&	2001/04/14 17:54&	--&	--&	--&	--&	--&	--&	*\\
51&	2001/04/26 12:30&	09433&	N17W31&	2001/04/26 08:30&	2001/04/25 14:06&	--&	--&	--&	--&	--&	--&	--&	--&	\\
52&	2001/05/10 01:31&	09445&	N25W80&	--&	--&	--&	--&	--&	--&	--&	--&	--&	--&	\\
53&	2001/07/19 10:30&	09537&	S08W62&	--&	--&	--&	--&	--&	--&	--&	--&	--&	--&	\\
54&	2001/09/15 11:54&	09608&	S21W49&	2001/09/14 16:31&	--&	--&	--&	--&	--&	--&	--&	--&	--&	*\\
55&	2001/10/01 05:30&	09628&	S20W84&	2001/10/01 01:54&	2001/09/30 10:34&	--&	--&	--&	--&	--&	--&	--&	--&	*\\
56&	2001/10/19 16:50&	09661&	N15W29&	2001/10/19 15:06&	2001/10/19 11:50&	2001/10/19 01:27&	--&	--&	--&	--&	--&	--&	--&	*\\
57&	2001/10/25 15:26&	09672&	S16W21&	--&	--&	--&	--&	--&	--&	--&	--&	--&	--&	\\
58&	2001/11/04 16:35&	09684&	N06W18&	2001/11/04 14:40&	2001/11/04 14:10&	2001/11/04 13:20&	2001/11/04 11:55&	--&	--&	--&	--&	--&	--&	*\\
59&	2001/11/22 20:30&	09698&	S24W68&	--&	--&	--&	--&	--&	--&	--&	--&	--&	--&	*\\
60&	2001/11/22 23:30&	09704&	S15W34&	2001/11/22 21:30&	--&	--&	--&	--&	--&	--&	--&	--&	--&	*\\
61&	2001/12/18 18:30&	09739&	S13W71&	2001/12/18 16:54&	2001/12/18 03:06&	--&	--&	--&	--&	--&	--&	--&	--&	\\
62&	2001/12/26 05:30&	09742&	N08W54&	2001/12/26 02:06&	--&	--&	--&	--&	--&	--&	--&	--&	--&	*\\
63&	2002/01/14 05:35&	fp&	S28W83&	--&	--&	--&	--&	--&	--&	--&	--&	--&	--&	*\\
64&	2002/02/20 06:30&	09825&	N12W72&	2002/02/20 03:30&	--&	--&	--&	--&	--&	--&	--&	--&	--&	*\\
65&	2002/03/15 23:06&	09866&	S08W03&	--&	--&	--&	--&	--&	--&	--&	--&	--&	--&	*\\
  \hline
\end{tabular}
\end{table}
\end{landscape}

\begin{landscape}
\begin{table}[!htbp]
{\bf Table~\ref{table.1}} (Continued)
\linespread{1.1}
\setlength{\tabcolsep}{1.1pt}
\tiny
\centering
\begin{tabular}{cccccccccccc>{\centering}p{1.8cm}>{\centering}p{1.8cm}c}
  \hline
  \hline
\multirow{2}{*}{No.}&\multirow{2}{1.8cm}{\centering{Onset time of main CME$^a$}} &
\multirow{2}{*}{AR$^b$}&\multirow{2}{*}{Loc.$^c$}&\multicolumn{10}{c}{Onset time of the preceding CMEs$^d$}&
\multirow{2}{*}{Comm.$^e$}\\
\cline{5-14}
&&&&p1&p2&p3&p4&p5&p6&p7&p8&p9&p10&\\
\hline
(1)&(2)&(3)&(4)&(5)&(6)&(7)&(8)&(9)&(10)&(11)&(12)&(13)&(14)&(15)\\
\hline

66&	2002/03/18 02:54&	09866&	S09W46&	2002/03/17 20:06&	--&	--&	--&	--&	--&	--&	--&	--&	--&	*\\
67&	2002/03/20 23:54&	09870&	S20W60&	2002/03/20 21:08&	2002/03/20 17:54&	2002/03/20 10:06&	--&	--&	--&	--&	--&	--&	--&	\\
68&	2002/04/17 08:26&	09906&	S14W34&	2002/04/16 11:06&	--&	--&	--&	--&	--&	--&	--&	--&	--&	*\\
69&	2002/04/21 01:27&	09906&	S14W84&	2002/04/20 16:50&	--&	--&	--&	--&	--&	--&	--&	--&	--&	*\\
70&	2002/04/30 23:26&	09914&	N06W78&	2002/04/30 19:06&	2002/04/30 16:11&	2002/04/30 10:26&	2002/04/30 09:50&	2002/04/30 08:06&	2002/04/30 02:26&	--&	--&	--&	--&	\\
71&	2002/05/07 00:06&	09929&	N22W66&	2002/05/06 08:26&	2002/05/06 06:06&	--&	--&	--&	--&	--&	--&	--&	--&	\\
72&	2002/05/22 00:06&	09948&	S15W70&	2002/05/21 16:06&	2002/05/21 15:26&	--&	--&	--&	--&	--&	--&	--&	--&	\\
73&	2002/05/22 03:50&	09948&	S15W70&	2002/05/22 00:06&	--&	--&	--&	--&	--&	--&	--&	--&	--&	*\\
74&	2002/07/15 21:30&	10030&	N19W01&	2002/07/15 20:30&	2002/07/15 14:06&	2002/07/15 11:30&	2002/07/15 00:30&	--&	--&	--&	--&	--&	--&	*\\
75&	2002/07/18 08:06&	10030&	N19W30&	--&	--&	--&	--&	--&	--&	--&	--&	--&	--&	\\
76&	2002/08/03 19:31&	10039&	S16W76&	2002/08/03 17:30&	2002/08/03 13:31&	2002/08/03 12:30&	2002/08/02 19:31&	--&	--&	--&	--&	--&	--&	\\
77&	2002/08/06 18:25&	fp&	S38W18&	--&	--&	--&	--&	--&	--&	--&	--&	--&	--&	\\
78&	2002/08/14 02:30&	10061&	N09W54&	--&	--&	--&	--&	--&	--&	--&	--&	--&	--&	*\\
79&	2002/08/16 06:06&	10061&	N07W83&	2002/08/16 05:30&	2002/08/16 01:54&	2002/08/15 22:30&	2002/08/15 18:06&	2002/08/15 15:30&	--&	--&	--&	--&	--&	\\
80&	2002/08/20 08:54&	10069&	S10W38&	2002/08/20 04:06&	2002/08/20 01:54&	2002/08/19 23:03&	2002/08/19 15:03&	2002/08/19 11:06&	--&	--&	--&	--&	--&	\\
81&	2002/08/22 02:06&	10069&	N07W62&	2002/08/21 17:30&	2002/08/21 11:30&	2002/08/21 07:31&	2002/08/21 03:54&	--&	--&	--&	--&	--&	--&	*\\
82&	2002/08/24 01:27&	10069&	N02W81&	2002/08/23 20:50&	2002/08/23 20:06&	--&	--&	--&	--&	--&	--&	--&	--&	*\\
83&	2002/09/06 02:06&	10095&	N08W31&	2002/09/05 23:30&	2002/09/05 21:56&	2002/09/05 14:30&	--&	--&	--&	--&	--&	--&	--&	*\\
84&	2002/11/09 13:31&	10180&	S12W29&	2002/11/09 10:56&	2002/11/09 09:54&	2002/11/08 18:30&	2002/11/08 17:30&	--&	--&	--&	--&	--&	--&	*\\
85&	2002/11/10 03:30&	10180&	S12W37&	2002/11/09 13:31&	2002/11/09 10:56&	2002/11/09 09:54&	--&	--&	--&	--&	--&	--&	--&	\\
86&	2002/11/11 15:52&	10180&	S07W59&	2002/11/11 07:54&	2002/11/11 02:30&	--&	--&	--&	--&	--&	--&	--&	--&	\\
87&	2002/12/19 22:06&	10299&	N15W09&	2002/12/19 08:06&	--&	--&	--&	--&	--&	--&	--&	--&	--&	\\
88&	2002/12/22 03:30&	10223&	N23W42&	2002/12/21 05:54&	--&	--&	--&	--&	--&	--&	--&	--&	--&	\\
89&	2003/01/27 22:23&	10267&	S17W23&	--&	--&	--&	--&	--&	--&	--&	--&	--&	--&	\\
90&	2003/03/17 19:54&	10134&	S14W39&	--&	--&	--&	--&	--&	--&	--&	--&	--&	--&	\\
91&	2003/03/18 12:30&	10314&	S15W46&	2003/03/18 10:30&	2003/03/18 07:31&	2003/03/18 06:30&	2003/03/17 19:54&	--&	--&	--&	--&	--&	--&	\\
92&	2003/03/19 02:30&	10314&	S16W56&	2003/03/19 01:32&	2003/03/18 14:54&	2003/03/18 13:54&	--&	--&	--&	--&	--&	--&	--&	\\
93&	2003/05/28 00:50&	10365&	S07W20&	2003/05/27 23:50&	2003/05/27 22:06&	2003/05/27 17:26&	2003/05/27 06:50&	--&	--&	--&	--&	--&	--&	*\\
94&	2003/05/31 02:30&	10365&	S07W65&	2003/05/31 00:30&	2003/05/30 06:26&	--&	--&	--&	--&	--&	--&	--&	--&	*\\
95&	2003/10/26 17:54&	10484&	N04W43&	2003/10/26 05:30&	2003/10/26 01:31&	--&	--&	--&	--&	--&	--&	--&	--&	*\\
96&	2003/10/27 08:30&	10484&	N03W48&	2003/10/27 04:30&	2003/10/27 04:06&	2003/10/26 22:30&	2003/10/26 20:18&	2003/10/26 19:54&	2003/10/26 17:54&	--&	--&	--&	--&	\\
97&	2003/10/29 20:54&	10486&	S15W02&	2003/10/29 10:16&	--&	--&	--&	--&	--&	--&	--&	--&	--&	*\\
98&	2003/11/02 09:30&	10486&	S17W55&	2003/11/02 08:54&	2003/11/02 08:30&	2003/11/02 06:30&	2003/11/02 02:06&	2003/11/01 23:06&	2003/11/01 14:54&	--&	--&	--&	--&	*\\
99&	2003/11/02 17:30&	10486&	S17W56&	2003/11/02 11:30&	2003/11/02 09:30&	2003/11/02 08:54&	2003/11/02 08:30&	2003/11/02 06:30&	2003/11/02 02:06&	2003/11/01 23:06&	--&	--&	--&	*\\
100&	2003/11/03 10:05&	10488&	N08W77&	2003/11/03 01:59&	--&	--&	--&	--&	--&	--&	--&	--&	--&	\\
101&	2003/11/04 19:54&	10486&	S19W83&	2003/11/04 12:54&	2003/11/04 12:06&	2003/11/04 04:54&	--&	--&	--&	--&	--&	--&	--&	*\\
102&	2003/11/07 15:54&	10495&	S21W89&	2003/11/06 17:30&	--&	--&	--&	--&	--&	--&	--&	--&	--&	\\
103&	2003/11/11 13:54&	10498&	S04W63&	2003/11/10 23:54&	--&	--&	--&	--&	--&	--&	--&	--&	--&	\\
104&	2003/11/20 08:06&	10501&	N01W08&	2003/11/20 02:50&	2003/11/19 15:06&	2003/11/19 14:06&	--&	--&	--&	--&	--&	--&	--&	*\\
105&	2003/12/02 10:50&	10508&	S19W89&	2003/12/02 05:50&	--&	--&	--&	--&	--&	--&	--&	--&	--&	*\\
106&	2004/04/08 10:30&	10588&	S15W11&	--&	--&	--&	--&	--&	--&	--&	--&	--&	--&	\\
107&	2004/04/11 04:30&	10588&	S16W46&	2004/04/10 18:54&	2004/04/10 17:54&	2004/04/10 08:06&	--&	--&	--&	--&	--&	--&	--&	*\\
108&	2004/07/25 14:54&	10652&	N08W33&	2004/07/25 13:31&	2004/07/25 11:54&	2004/07/24 23:54&	2004/07/24 21:30&	--&	--&	--&	--&	--&	--&	*\\
109&	2004/11/07 16:54&	10696&	N09W17&	2004/11/07 09:54&	2004/11/07 09:06&	2004/11/07 03:18&	2004/11/07 01:54&	2004/11/06 23:54&	--&	--&	--&	--&	--&	*\\
110&	2004/11/09 17:26&	10696&	N08W51&	--&	--&	--&	--&	--&	--&	--&	--&	--&	--&	\\
111&	2004/11/10 02:26&	10696&	N09W49&	2004/11/09 17:26&	--&	--&	--&	--&	--&	--&	--&	--&	--&	*\\
112&	2004/12/03 00:26&	10708&	N08W02&	--&	--&	--&	--&	--&	--&	--&	--&	--&	--&	\\
113&	2005/01/04 09:30&	fp&	N15W60&	--&	--&	--&	--&	--&	--&	--&	--&	--&	--&	\\
114&	2005/01/15 23:06&	10720&	N15W05&	2005/01/15 14:54&	2005/01/15 07:31&	2005/01/15 06:30&	--&	--&	--&	--&	--&	--&	--&	*\\
115&	2005/01/17 09:54&	10720&	N15W25&	2005/01/17 09:30&	2005/01/16 20:30&	2005/01/16 18:30&	--&	--&	--&	--&	--&	--&	--&	*\\
116&	2005/01/19 08:29&	10720&	N15W51&	2005/01/18 23:08&	--&	--&	--&	--&	--&	--&	--&	--&	--&	\\
117&	2005/01/20 06:54&	10720&	N14W61&	2005/01/20 04:06&	2005/01/19 23:42&	2005/01/19 19:31&	2005/01/19 10:54&	2005/01/19 08:29&	--&	--&	--&	--&	--&	*\\
118&	2005/02/17 00:06&	10734&	S03W24&	--&	--&	--&	--&	--&	--&	--&	--&	--&	--&	\\
119&	2005/05/06 03:30&	10756&	S04W71&	2005/05/06 02:30&	2005/05/05 21:54&	2005/05/05 20:58&	2005/05/05 09:54&	--&	--&	--&	--&	--&	--&	\\
120&	2005/05/06 11:54&	10756&	S04W76&	2005/05/06 10:30&	2005/05/06 03:30&	--&	--&	--&	--&	--&	--&	--&	--&	\\
121&	2005/07/09 22:30&	10786&	N12W28&	--&	--&	--&	--&	--&	--&	--&	--&	--&	--&	\\
122&	2005/07/13 14:30&	10786&	N08W79&	2005/07/13 12:54&	2005/07/13 03:06&	2005/07/13 02:30&	2005/07/12 23:30&	2005/07/12 18:06&	2005/07/12 16:54&	2005/07/12 16:30&	--&	--&	--&	*\\
123&	2005/08/22 01:31&	10798&	S11W54&	2005/08/21 12:06&	--&	--&	--&	--&	--&	--&	--&	--&	--&	\\
124&	2005/08/22 17:30&	10798&	S12W60&	2005/08/22 14:18&	2005/08/22 09:06&	2005/08/22 05:30&	2005/08/22 03:30&	2005/08/22 01:31&	--&	--&	--&	--&	--&	*\\
125&	2006/12/13 02:54&	10930&	S06W23&	2006/12/12 20:28&	--&	--&	--&	--&	--&	--&	--&	--&	--&	*\\
126&	2006/12/14 22:30&	10930&	N06W46&	2006/12/14 20:30&	2006/12/13 23:48&	--&	--&	--&	--&	--&	--&	--&	--&	*\\
  \hline
\end{tabular}
\begin{itemize}
\tiny
  \item[$^a$] first appearance time of the main CME in LASCO/C2.
  \item[$^b$] NOAA active source region (AR) number, `fp' denotes no NOAA active region number and possibly a filament/prominence eruption.
  \item[$^c$] location of source region.
  \item[$^d$] first appearance time of the preceding CME of the main CME within 24hr. `p1' column shows the onset time of the closest preceding CME, `p2' shows the onset time of the second closest preceding CME, and so on. '-' denotes that no preceding CME is identified.
  \item[$^e$] `*' denotes the event where the main CME produces a large SEP event.
\end{itemize}
\end{table}
\end{landscape}

\end{document}